\documentclass[preprint2]{aastex}

\bibpunct{(}{)}{,}{}{,}{,}

\shorttitle{Propagation of MHD kink waves}
\shortauthors{Morton et al.}

\begin{document}
\title{The generation and damping of propagating MHD kink waves in the solar atmosphere }

\author{R. J. Morton\altaffilmark{1}} 

\affil{Mathematics and Information Sciences, Northumbria University, Newcastle Upon Tyne,
NE1 8ST, UK}
\email{richard.morton@northumbria.ac.uk}

\author{G. Verth\altaffilmark{2}, }
\affil{Solar Physics and Space Plasma Research Centre
(SP$^2$RC), The University of Sheffield, Hicks Building, Hounsfield
Road, Sheffield S3 7RH, UK}
\email{g.verth@sheffield.ac.uk}

\author{A. Hillier \altaffilmark{3}} 

\affil{Kwasan and Hida Observatories, Kyoto University, 17 Ohmine-cho Kita Kazan, Yamashina-ku, Kyoto City, Kyoto 607-8471, Japan}

\author{R. Erd\'elyi\altaffilmark{2}}
\email{robertus@sheffield.ac.uk}

\date{Received /Accepted}
\begin{abstract}
The source of the non-thermal energy required for the heating of the upper solar atmosphere to 
temperatures in excess of a million degrees and the acceleration of the solar wind to hundreds of 
kilometres per second is still unclear. One such mechanism for providing the required energy flux is 
incompressible torsional Alfv\'en and kink magnetohydrodynamic (MHD) waves, which are magnetically 
dominated waves supported by the Sun's pervasive and complex magnetic field. In particular, 
propagating MHD kink waves have recently been observed to be ubiquitous throughout the solar 
atmosphere, but, until now, critical details of the transport of the kink wave energy throughout the Sun's 
atmosphere were lacking. Here, the ubiquity of the waves is exploited for statistical studies in the highly 
dynamic solar chromosphere. This large-scale investigation allows for the determination of the 
chromospheric kink wave velocity power spectra, a missing link necessary for determining the energy 
transport between the photosphere and corona. Crucially, the power spectra contains evidence for 
horizontal photospheric motions being an important mechanism for kink wave generation in the quiescent 
Sun. In addition, a comparison to measured coronal power spectra is provided for the first time, revealing 
frequency-dependent transmission profiles suggesting there is enhanced damping of kink waves in the 
lower corona.
\end{abstract}

\keywords{Sun: Photosphere, Sun: Chromosphere, Sun: Corona, Waves, MHD}

\section{Introduction}
It has long been proposed that the kinetic energy in stellar convective envelopes is transferred 
throughout stellar atmosphere by magnetic fields (\citealp{OST1961}; \citealp{KUPetal1981}; 
\citealp{NARULM1996};
\citealp{KLI2006}; \citealp{DEPetal2007};
\citealp{TOMetal2007}; \citealp{JESetal2009}; \citealp{MCIetal2011}; \citealp{WEDetal2012}; 
\citealp{MORetal2012c}). A number of popular theories assume that the horizontal components of the 
motion of convective granulation, observed at the solar photospheric surface, excite incompressible MHD 
transversal waves in magnetic flux concentrations (\citealp{KUPetal1981}; \citealp{NARULM1996}; 
\citealp{KLI2006}). These incompressible motions can either be perpendicular to the constant magnetic 
surfaces (i.e. kink modes) or tangential to these surfaces (i.e. torsional Alfv\'en 
modes) (\citealp{ERDFED2007}). More recently, advanced analytical and numerical models have used 
either theoretical turbulent convective spectra (\citealp{MUSULM2002}; \citealp{FAWetal2002}) or velocity 
power spectra estimated from observations of the solar granulation (\citealp{RUDetal1997}; 
\citealp{CRAVAN2005}; \citealp{CRAetal2007}; \citealp{CHOetal1993, CHOetal1993b}; 
\citealp{MATSHI2010}; \citealp{ANTSHI2010}) as their input spectrum for generating incompressible 
waves in stellar atmospheres. These models have had some success in generating the necessary non-
thermal energy needed for plasma heating in the atmospheric layers and providing the necessary energy 
flux for accelerating solar winds. However, it was not clear whether the velocity power spectra derived 
from the horizontal motions were the physically appropriate input for models. This was in part due to a 
number of recent observations that demonstrated different incompressible wave excitation mechanisms, 
e.g., vortices (\citealp{WEDetal2012}; \citealp{MORetal2013}),
magnetic reconnection (\citealp{Heetal2009}). However, previously the main 
restriction was the dearth of large-scale observational studies of wave behaviour in the various solar 
atmospheric layers, with which the outputs of these numerical models could be compared.


\begin{figure*}[!htp]
\centering
\includegraphics[scale=0.8, clip=true, viewport=0.0cm 0.0cm 17.5cm 15.0cm]{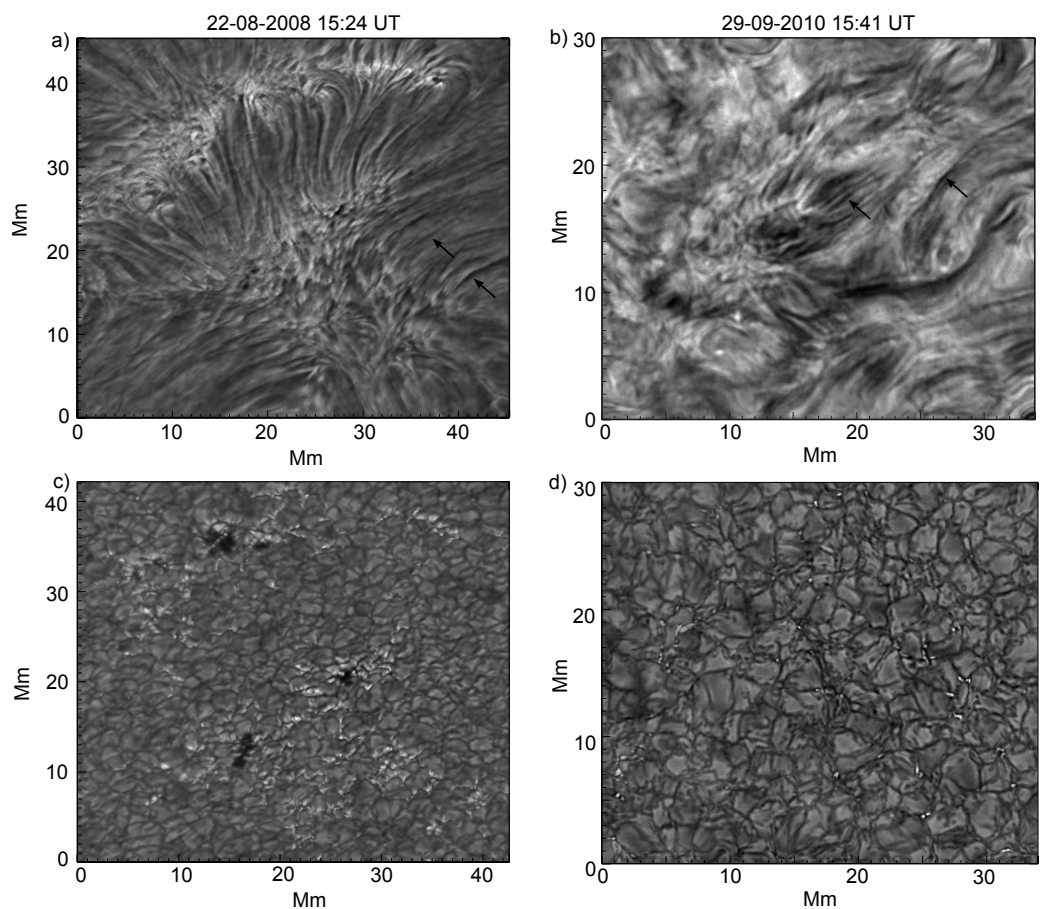} 
\caption{The solar atmosphere observed by ROSA. (a) The solar chromosphere in a magnetically active 
region as seen with an H-alpha filter. The image displays a 45 Mm by 45 Mm sub-region for D1. (b) H-
alpha image of the quiescent solar chromosphere, showing a 34 Mm by 31 Mm sub-region for D2. The 
existence of fine-scale structuring in the bandpass is evident in both datasets, with both 
spicules/mottles and cell-spanning fibrils identifiable. {Examples of fibrils are highlighted by the black arrows.} 
(c) Corresponding G-band image for D1, which 
reveals magnetic bright points, larger magnetic pores and the solar granulation. (d) The corresponding 
G-band image for D2, which shows only the magnetic bright points and the granulation. The G-band 
images depict the solar photosphere that lies directly under the H-alpha 
chromosphere.}\label{fig:rosafov}
\end{figure*}

Understanding the role of MHD wave dynamics in solar plasma heating is crucial, but, until recently, 
detailed studies have proved challenging. It is only in the last few years that space- and ground-based 
imaging instruments have achieved the necessary spatial resolution to resolve the incompressible 
motions of the fine-scale magnetic structure. This has resulted in a wealth of evidence from a wide range 
of instruments demonstrating that incompressible wave energy is ubiquitous throughout the 
chromosphere (\citealp{DEPetal2007}; \citealp{MORetal2012c}; \citealp{PERetal2012}, transition region 
(\citealp{MCIetal2011}) and corona (\citealp{TOMetal2007}; \citealp{MCIetal2011}). MHD kink wave 
energy estimates from observations hint that the chromospheric fine structure exhibits much more 
energetic motion than the coronal fine structure (\citealp{MCIetal2011}; \citealp{DEPetal2007}; 
\citealp{TOMetal2007}; \citealp{MORetal2012c}). However, no attempt has yet been made to demonstrate 
the transport of kink wave energy between the different atmospheric layers, which is essential for 
distinguishing between various heating models. Again, this is in part due to the limited nature of 
previous wave studies. Here, we provide a major missing link in this problem by determining 
chromospheric velocity power spectra from observations. This allows for the comparison of the chromospheric power
spectra to other velocity power spectra derived at different altitudes in the solar atmosphere. The comparisons 
reveal the first observational details of kink energy transport through the solar atmosphere.

\begin{figure*}[!t]
\centering
\includegraphics[scale=0.9, clip=true, viewport=0.0cm 0.0cm 15.5cm 11.5cm]{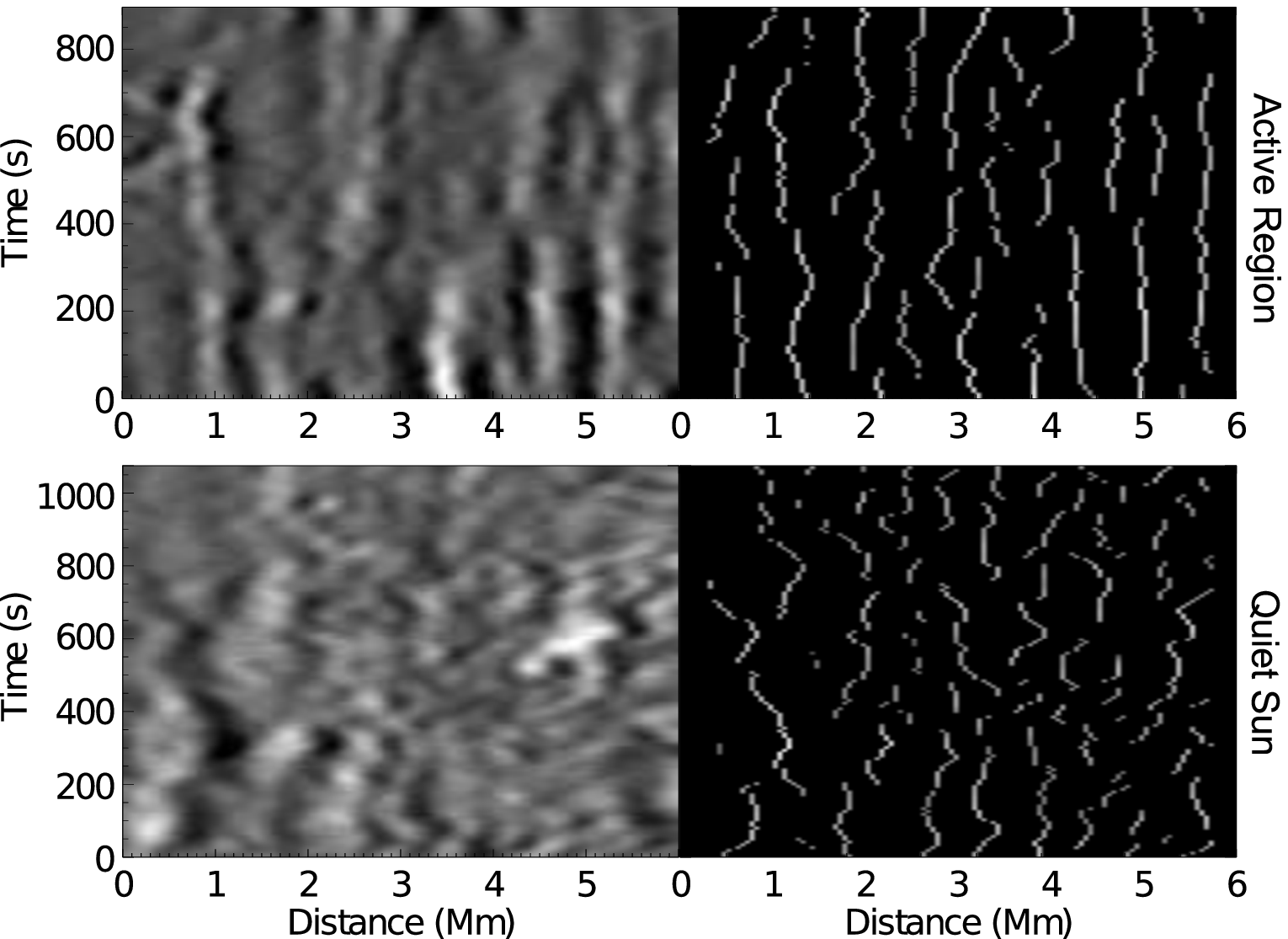}
\caption{The left panels show example unsharp masked time-distance diagrams from the active and quiet Sun fibrils. The right hand panels are the results of the feature tracking algorithm that reveal the displacements of the fibrils.  }\label{fig:track}
\end{figure*}

\section{Observations}

The hydrogen alpha (H-alpha) spectral line has proved invaluable for exploration of the magnetically 
dominated chromosphere (\citealp{RUT2012}), in particular for incompressible wave studies. Here we use 
two H-alpha datasets taken with Rapid Oscillations in the Solar Atmosphere (ROSA - 
\citealp{JESetal2010}) at the Dunn Solar Telescope at Sacramento Peak, USA. Both datasets are positioned 
relatively close to disk centre, which implies the line-of-sight (LOS) is almost vertically down into the 
solar atmosphere.

 The datasets were obtained at 15:24-16:35~UT on 22 August 2008 (D1) and 15:41-16:51~UT on 29 
September 2010 (D2).  The first dataset (D1) is a magnetically active region located at (N14.8, E40.2). The 
second dataset (D2) is a $69''.3\times69''.1$ region of the quiescent solar atmosphere, positioned close 
to disk centre (N0.9, W6.8). Both datasets were taken with a spatial sampling of 0''.069 pixel$^{-1}$. 
During the observations, high-order adaptive optics (\citealp{RIM2004}) were used to correct for 
wave-front deformations in real time.

The H$\alpha$ data was sampled at 10 frames s$^{-1}$ for D1 and 2.075 frames s$^{-1}$ for D2. All images were improved by using 
speckle reconstruction (\citealp{WOEetal2008}). The cadence of the two time-series are
 6.4~s for D1 and 7.7~s for D2. To ensure accurate co-alignment in all bandpasses, the
broadband times series were Fourier co-registered and de-stretched (\citealp{JESetal2007}).

D1 focuses on a region of relatively strong magnetic activity, with the G-band images 
of the photosphere (Figure~\ref{fig:rosafov}) revealing underlying small ($\sim$200~km diameter - 
\citealp{CROetal2010}) bright, intense magnetic elements (\citealp{BERetal2001}) located in the 
intergranular network and larger (1000~km) dark magnetic pores. In contrast, the G-band for the second 
dataset (D2) reveals only magnetic bright points, suggesting the total magnetic flux underlying the D2 
H-alpha region is significantly less than in D1. The affect of the differing magnetic fluxes is reflected in 
the chromospheric features detected in H-alpha. {Here, we are mainly interested in the dark fine structure known as fibrils. 
Modelling of H$\alpha$ line formation suggests that the line core intensity is inversely proportional to density, implying that the fibrils are over dense compared to their ambient environment (\citealp{LEEetal2012}). A few of these features are highlighted in Figure~\ref{fig:rosafov}.} D1 has very clear, distinct and 
ordered fibrillar structures that are long lasting. The fine-scale structure in D2 is, however, less distinct and is not so 
ordered. However, regions of elongated fibrillar structures that outline the chromospheric magnetic field 
(\citealp{LEEetal2012}) can still be identified. The foot-points of the fibrils appear to be rooted in the 
regions of intense magnetic field (\citealp{REAetal2012}), with the other foot-points in intergranular 
lanes.

\vspace{0.2cm}

In Section~4 we make use of results obtained from different instruments. We provide here brief details of 
the datasets used.

\begin{figure}[!htp]
\centering
\includegraphics[scale=0.85, clip=true, viewport=0.0cm 0.0cm 10.cm 9.5cm]{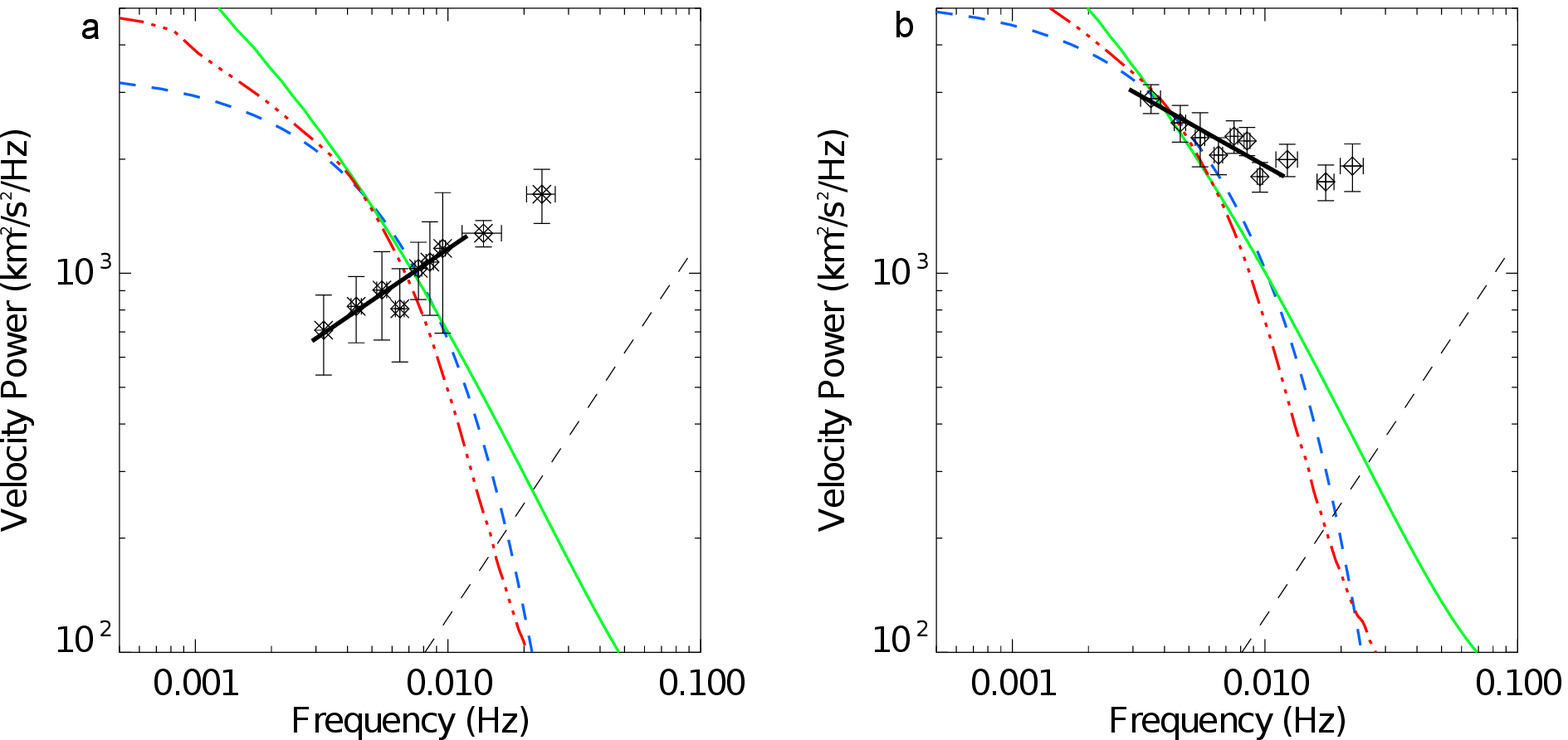} 
\includegraphics[scale=0.85, clip=true, viewport=9.9cm 0.0cm 20.cm 9.5cm]{fig5_b.eps} 
\caption{Comparing the power spectra of the photosphere and chromosphere. (a) The figure displays the 
median velocity power in the chromosphere as a function of frequency for the magnetically active region 
D1. The velocity power data points are calculated from frequency binned PDFs and the vertical error 
bars show the standard deviation of the velocity power in the bins. Over-plotted are the photospheric 
velocity power spectra of 
horizontal motions; \citealp{MATKIT2010}: red – dash dot, and \citealp{CHITetal2012}: green 
solid and blue dash. The photospheric data has been scaled by a constant factor for comparison. (b) 
Same as (a) but for the quiescent Sun region D2. 
}\label{fig:gen}
\end{figure}

The photospheric velocity power spectrum (red-dot dashed line Figure~\ref{fig:gen}) was obtained {by measuring the motions of 
granulation} using the Hinode Solar Optical Telescope (SOT) G-band channel (further details are given in 
\citealp{MATKIT2010}). The data were obtained on the 18 March 2007 at 07:56~UT. The photospheric 
velocity power spectrum derived from the data is representative of the quiescent Sun and is comparable 
to power spectra derived from 13 other datasets.

 The other two photospheric power spectra (blue dashed and green-solid lines Figure~\ref{fig:gen}) were 
obtained {by measuring the motions of individual magnetic bright points} using the Swedish Solar Telescope (further details are 
given in \citealp{CHITetal2012}). The data were obtained on the 18 June 2006 from 13:10~UT.

The coronal velocity power spectrum was obtained using the Coronal Multi-channel Polarimeter (CoMP) 
and details are given in \cite{TOMMCI2009}. The data were obtained on the 30 October 2005.

\begin{figure*}[!t]
\centering
\includegraphics[scale=1.0, clip=true, viewport=0.0cm 0.0cm 17.5cm 5.7cm]{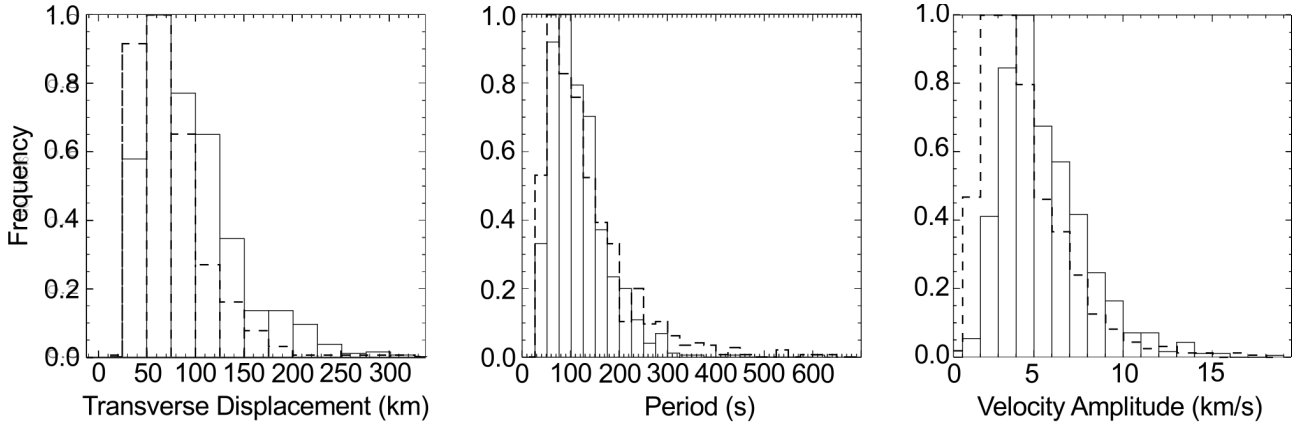}
\caption{
Histograms of measured properties of transverse motions for chromospheric fine structure. The 
histograms show, from left to right, transverse displacement amplitude, the period and transverse 
velocity amplitude. The dashed and solid lines correspond to the results from D1 and D2, respectively. 
The mean and standard deviations are $P_{D1}=130\pm92$~s, $A_{D1}=73\pm36$~km, 
$v_{D1}=4.4\pm2.4$~km\,s$^{-1}$ and $P_{D2}=116\pm59$~s, $A_{D2}=94\pm47$~km, 
$v_{D2}=5.5\pm2.4$~km\,s$^{-1}$. }\label{fig:hist}
\end{figure*}

\section{Data analysis}
On analysing the H-alpha movies of these two regions, the dynamic behaviour of the chromospheric fine 
structure is evident. Our interest is directed towards the axial transverse displacement of the 
chromospheric fibrils, which is the unique signature of MHD kink wave motion. The transverse fibril 
displacements are identified and measured using a semi-automated tracking mechanism 
(\citealp{MORetal2013}). We give a brief description here.

{The first step in the analysis of the observed transverse waves is to apply an unsharp mask procedure to the H$\alpha$ 
images. Cross-cuts are placed perpendicular to the axis of the fibrils and time-distance plots are created, examples of which are shown 
in Figure~\ref{fig:track}. The time-distance plots clearly reveal the transverse motions of the fibril structures. For each fibril segment in 
each time-slice, the central pixel is identified and provides the fibril tracks through time. The result is shown in the right hand panels of 
Fig.~\ref{fig:track}. }

{
It can be seen that the observed motions are sinusoidal in nature, with one period typically dominating the motion of an individual fibril 
at a specific time. This mono-periodic behaviour is typical of chromospheric transverse waves, although the measurement of many such features at various times and numerous locations shows a rather broad range of frequencies (e.g., 
\citealp{OKADEP2011}; \citealp{MORetal2013}).}
  
{ 
The fibril tracks are then fitted with a Levenberg-Marquardt non-linear fitting algorithm (mpfit.pro  - \citealp{MAR2009}). A function of 
the form \begin{equation}
F(t)=G(t)+A\sin(\omega t-\phi)
\end{equation}
is used to fit the oscillations. Here, $G(t)$ is a linear function, $A$ is the displacement amplitude of the oscillation, 
$\omega$ the frequency and $\phi$ the phase of the oscillation. The fitting algorithm is supplied with the errors on 
each data point, where it is assumed the given error is the one-$\sigma$ uncertainty. The fitting of a sinusoidal function of a single 
frequency leads to the measurement of the dominant oscillation frequency of the fibril. Residuals exist between the fit and the 
measured track, which could be the signature of a superposition of different frequencies.} 

The analysed fibrils cover the entire field of view for both datasets. In D1 and 
D2, a total of 744 and 841 sinusoidal transverse displacements are measured, respectively. Histograms 
of the periods, transverse amplitudes and velocity amplitudes are given in Figure~\ref{fig:hist}. The 
previous observations in fibrils (\citealp{MORetal2012c, MORetal2013}; \citealp{KURetal2012}), off-limb 
spicules (\citealp{DEPetal2007}; \citealp{PERetal2012}) and other small-scale chromospheric features 
(\citealp{SEKetal2012}) are found to be consistent with our extensive statistical study here. Note that 
longer-period waves are observable in magnetically more active regions than in the quiescent Sun. This is 
unlikely to be physical but a consequence that the visible lifetimes of the fibrillar structures are 
longer in the active region than in quiescent regions.

\begin{figure*}[!htp]
\centering
\includegraphics[scale=0.9, clip=true, viewport=0.0cm 0.0cm 16.cm 14.5cm]{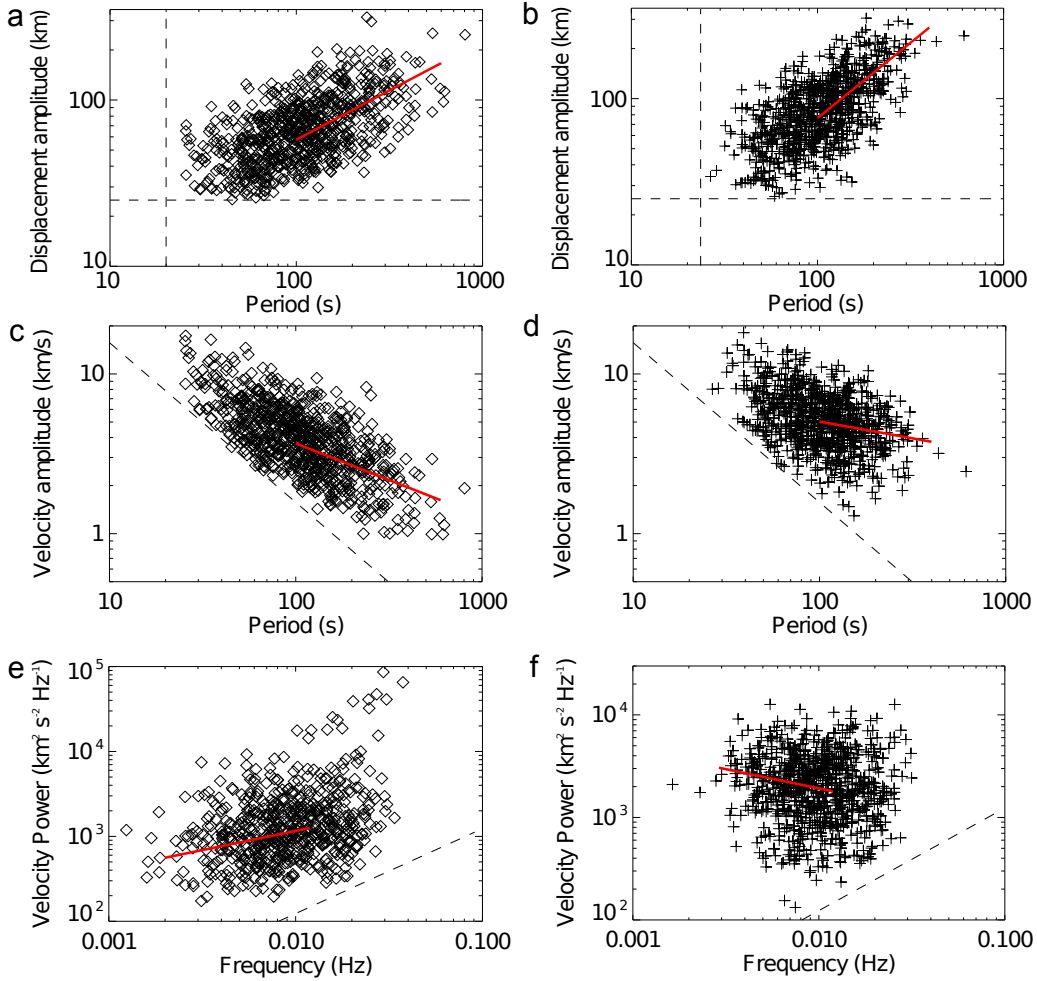} 
\caption{Observed wave properties as functions of period. Scatter plots of the transverse displacement 
versus period (a, b), velocity amplitude versus period (c, d) and the chromospheric power spectrum (e, f) for 
low-frequency kink waves. The diamonds correspond to the results from the dataset D1 and the 
crosses are the results from D2. The blue dashed-dot and solid lines correspond to the 
weighted power law fits to the results from D1 and D2, respectively. The black dashed lines highlight the 
observational limitations. }\label{fig:vsf}
\end{figure*}

Next, we demonstrate how the observed transverse displacements (A) and velocity amplitudes (v) vary as a 
function of period (Figure~\ref{fig:vsf}). In Figure~\ref{fig:vsf}a, b, transverse displacement versus period is 
plotted. The ability to detect waves with certain periods and displacement amplitudes are also subject to 
observational constraints based on the resolution of the DST and analysis techniques. The fitting 
technique locates the centre of the structure to within one pixel and, hence, has an error of $\pm25$~km 
(0.5~pixel) on each point, i.e., assuming that the centre of the structure were shifted over half a pixel in 
either direction; the minimum would be located in the neighbouring pixel. The fitting algorithm is 
supplied with the errors on each data point, where it is assumed the given error is the one-$\sigma$ 
uncertainty. The fitting routine then calculates the one-$\sigma$ error to each fit parameter. 
Theoretically, we only need three data points to be able to resolve a sinusoidal displacement,
hence, the minimum resolvable period is 20 (24)~s for data D1(D2). These limitations correspond to the
dashed lines showing the minimum measurable amplitudes and period over-plotted in 
Figure~\ref{fig:vsf}a. The constraining lines suggest, for the shortest measurable periods, the 
measured distribution of transverse amplitudes is likely influenced by the observational limitations. This 
effect is reduced for the quiescent Sun data (D2) because the typical measured displacements amplitudes 
are larger in D2 than in the active region (D1) (see also Figure~\ref{fig:hist}). 

To provide a fit to the generated data points, we first note the transverse displacement shows a 
log-normal distribution if projecting the points onto the displacement axis. We bin the data in the 
frequency domain, with bins of width 0.001~Hz between 0.004-0.01~Hz and widths of 0.05~Hz between 
0.01-0.02~Hz. Data less than 0.004~Hz are placed in one bin and data above 0.02 Hz are placed in 
another. For each frequency bin we take the log of the transverse displacement and plot a probability 
density function (PDF). A Gaussian is fit to 
each PDF and the centroid gives the median log displacement value while the width provides the standard 
deviation. It should be borne in mind that the bin for $<$0.004~Hz contains about half the 
number of data points as the other bins, i.e., it is less reliable. Further, the data points in the bins for 
frequencies $>0.01$~Hz are subject to larger errors and suffer the increased influence of the 
observational constraint.

The data points generated from the PDFs  are then fitted with a power law of the form $10^aP^b$ 
(where P is the period of the wave) computed with a non-linear Levenberg-Marquardt algorithm 
(\citealp{MAR2009}). The data points are weighted by the standard deviation of each PDF divided by the 
square root of the number of elements in each distribution, i.e., the standard error. In in attempt to 
negate the influence of the observational constraint on the results, the fit to the data is only for results 
with periods greater than 100 s. The fit gives a=0.58$\pm$0.15, b=0.59$\pm$0.07 for D1 and 
a=0.07$\pm$0.12, b=0.91$\pm$0.05 for D2.

Maximum transverse velocity amplitudes for the kink waves can be obtained from the fit using the 
relation $v =\omega A$. For the minimum resolvable velocity amplitude we calculate the constraint as the minimum displacement
amplitude multiplied by $2\pi$ and divided by period. 


In Figure~\ref{fig:vsf} c, d, velocity amplitude versus period is displayed with on a log-log axis. 
In light of the observational constraints discussed above, the velocity amplitudes are also fit with a power 
law from P=100~s onwards. The measured power laws for D1 are a=1.48$\pm$0.16, 
b=-0.46$\pm$0.07 and for D2 are a=1.12$\pm$0.14, b=-0.21$\pm$0.06. The power of the 
observational constraint for both datasets is b=-1. 

As suggested by the fits to the transverse displacements, an increase in velocity amplitude for 
decreasing period is present. Therefore, this suggests that waves with higher frequency transport a greater amount of energy 
through the chromosphere than the waves with lower frequency. If this trend were to continue, it would give support to incompressible 
MHD wave heating theories in which higher frequency waves are the dominant source of the wave energy.

The next, and key, step in the analysis is to estimate the chromospheric velocity power density as a function of 
frequency (f=1/P) for kink waves (Figure~\ref{fig:vsf}e, f). {The one sided velocity power density, $W$, for a
time series $v(t)$ is given by
\begin{equation}\label{eq:vel_pow}
W=\frac{2(\Delta t)^2\mid V_n\mid^2}{T},
\end{equation}
\noindent where $V_n$ are the Fourier coefficients of $v(t)$, $\Delta t$ is the sample time and $T=N\Delta t$ is the length of the time series and $N$ is 
the number of samples. For a time series described by $v(t)=A\omega\sin(\omega t)$, it is found that
\begin{equation}\label{eq:four_comp}
\frac{\mid v_{max}\mid^2}{2}\approx2\frac{\mid V_{max}\mid^2}{N^2},
\end{equation}
where $\mid V_{max}\mid$ is the Fourier coefficient with the largest root mean square value.
The left hand side can be interpreted as the time averaged value of $v(t)$, i.e. 
$$
\frac{\mid v_{max}\mid}{\sqrt{2}}=\frac{\mid A\omega\mid}{\sqrt{2}},
$$ 
which for a velocity time series we denote $\langle v\rangle$.
Substituting Eq.~(\ref{eq:four_comp}) into (\ref{eq:vel_pow}) and using the relation $\Delta f=1/(N\Delta t)$, we obtain the following relation
for the velocity power
\begin{equation}
W=\frac{\langle v\rangle^2}{\Delta f}=\frac{v^2}{2\Delta f}.
\end{equation}\label{eq:vel_pow2}
The $\Delta f$ term acts as a scaling factor that is inversely proportional to the length of time the oscillatory signal is observed for.
For the two data sets used in this paper, the lifetime of the measured oscillations is proportional to $1/f$.
The observational constraint for the velocity power is then calculated using the form of Eq.~(\ref{eq:vel_pow2}), hence is given by the 
time averaged velocity constraint (i.e. $v/\sqrt{2}$) squared multiplied by $1/f$. }

The velocity power is plotted on log-log 
axis and is subject to a weighted fit of a power law function of the form $10^af^b$. Following the 
previous power law fits, the fit is calculated for $P>$100~s ($f<0.01$~Hz). The power law fit gives 
for D1 a=3.97$\pm$0.54, b=0.45$\pm$0.25 and for D2 a=2.53$\pm$0.23, b=-0.37$\pm$0.11.  The 
derived velocity power spectra provide us with a powerful tool that can be applied to compare velocity 
power spectra established in other layers of the solar atmosphere. Such a comparison can reveal details 
on the frequency-dependent transport of non-thermal energy through the solar atmosphere. 

{It should be noted that the given values for velocity power do not include the number density of measured events in 
frequency space. This is because the observed number of events at low and high frequencies is biased by the fibrils life-time and the 
limitations of the measurement technique, respectively. Scaling with respect to the number density would then introduce observational 
biases to the power spectra. In order to assess whether the power spectra is influenced by neglecting the number density, the power 
spectral density is calculated for the longest time-series via a Fourier transform. We find that the gradient of the obtained power spectra 
is in agreement with that obtained already. Hence, it appears neglecting the number density of events does not influence the power 
spectra.}

\section{Discussion and conclusions}
\subsection{Wave generation}
It is natural to wonder how MHD kink waves are actually generated. It may be expected that the wave’s 
spectra should contain a signature of their excitation mechanism, e.g., the horizontal photospheric 
motions. On the other hand, wave spectra can be modified due to mode conversion (e.g., \citealp{CARBOG2006}; 
\citealp{CALGOO2008}; \citealp{FEDetal2011}) at altitudes where the Alfv\'en speed equals the sound speed (thought to  be 
the low chromosphere in the quiet sun - e.g., \citealp{WEDetal2009}), due to reflection from strong gradients in plasma quantities 
present in the TR (e.g., \citealp{HOL1981};  \citealp{FEDetal2011}) {or non-linear processes associated with turbulent cascade 
(\citealp{CRAVAN2005}).}

{\cite{MORetal2013} raised the question whether current photospheric power spectra are a suitable input for numerical 
models of wave heating. The motions of the photospheric foot points of magnetic fields are typically measured from horizontal 
motions of 
magnetic bright points or granular motions. These motions are assumed to displace the whole flux element, with horizontal length 
scales proportional 
to the granulation ($10^3$~km) and on the time scales of granular flow patterns (a few minutes). It was suggested in 
\cite{VANBALLetal2011} that there may be the generation of additional transverse motions due to highly turbulent convective 
downflows (e.g., \citealp{VOGetal2005}). These transverse displacements would occur inside the magnetic elements on length scales 
shorter then the features ($\sim100$~km) and potentially have shorter associated time scales than the granular motions.}

 The data points from the generated PDFs are plotted 
in Figure~\ref{fig:gen}. In addition, the photospheric velocity power spectra, measured from two different 
characteristic sets  of quiescent photospheric phenomena {(granulation - \citealp{MATKIT2010}; magnetic bright points 
\citealp{CHITetal2012})}, are 
over-plotted. {It can be seen from these spectra that there is more power at the lower frequencies and the power drops of rapidly 
for time-scales less than a few minutes ($\sim5$~mHz).} The photospheric results are scaled up by factors of 20-70 for better 
visualisation. This is simply due to the smaller velocity amplitudes in the solar photosphere. The increase in amplitude from 
the photosphere to chromosphere is expected because of the decrease in {density} with height.

The gradient of the chromospheric power spectrum for D1 is relatively steep, showing an increase in velocity power with frequency. The
spectra does not appear to show a correlation with the photospheric power spectra. The lack of similarity could suggest that 
photospheric motions are not responsible for the driving of the waves in the active regions. Conversely, the photospheric 
velocity power spectra are derived for quiescent Sun regions.  Photospheric flow measurements 
(\citealp{TITetal1989}) show that flows are suppressed as the magnetic activity increases. This would potentially suppress the power at 
lower frequencies hence, the photospheric motions in active regions may produce waves with an alternate power spectra. To the best of 
our knowledge, there are no observational horizontal velocity power spectra for magnetically active regions to provide a comparison to. 

In contrast to the results for D1, a very good agreement exists between the gradients of the quiescent 
chromospheric (D2) and photospheric velocity power spectra for waves with $f<8$~mHz. The correlation indicates that the 
horizontal photospheric motions are potentially responsible for generating the low frequency chromospheric dynamics. It is worth 
noting that the power spectra for low-frequency transverse motions in prominences also displays a similar correlation with horizontal 
photospheric motions (\citealp{HILetal2013}). There is a flattening of the gradient of the velocity power spectra for frequencies 
$>8$~mHz (again seen in prominence wave power spectra). {At present, it is not possible to determine the extent to which this 
result is is influenced by observational artefacts (e.g., influence of the observational constraint) and how much additional power is due 
to genuine physical phenomena. The excess power could potentially 
be explained if there is an additional source of wave power at short spatial and temporal scales in the photosphere, e.g., turbulent 
convective downflows (\citealp{VANBALLetal2011}). }

 Further studies with a combination of higher cadences, improved spatial resolution and 
more advanced wave measurement routines are likely required to resolve this ambiguity. As noted in 
\cite{HILetal2013}, further work is required to establish a direct cause and effect relationship between the horizontal
motions and the transverse waves observed higher in the solar atmosphere, e.g., via the inclusion of phase spectra, which will be the aim of future studies.


\begin{figure}[!htp]
\centering
\includegraphics[scale=1.2, clip=true, viewport=0.0cm 0.0cm 7.02cm 10.5cm]{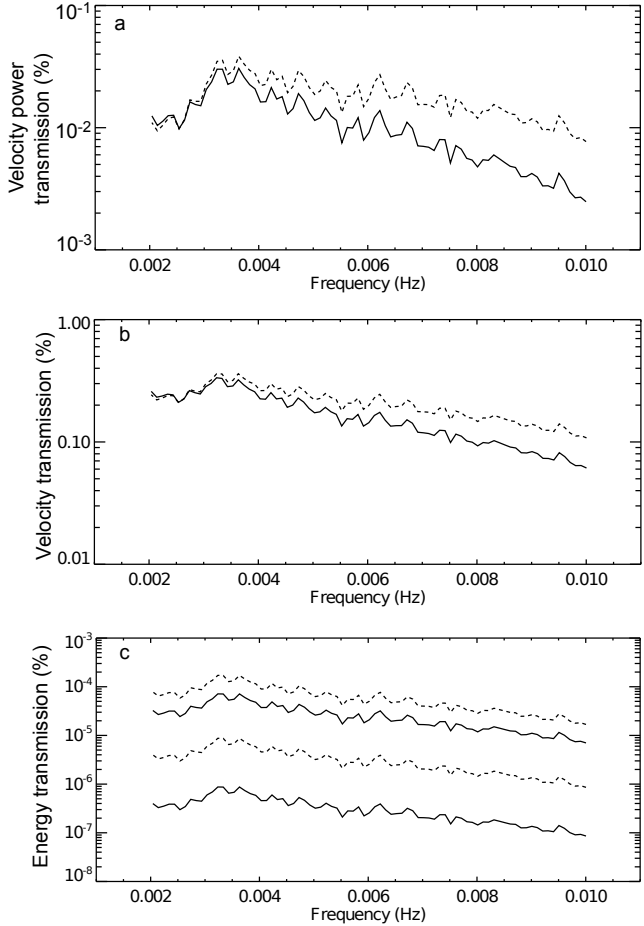} 
\caption{Transmission profiles from the chromosphere to the corona. A comparison of kink waves in the 
quiescent corona observed with CoMP and the quiescent chromosphere (D2). (a) The ratio of coronal 
velocity power to chromospheric velocity power. (b) The ratio of coronal velocity amplitude to 
chromospheric velocity amplitude. The solid lines are the ratios using the spatially averaged CoMP 
measurements. The dashed lines correspond to the ratio when using the estimated input 
power/velocity in the corona at a height of 20 Mm.  (c) The ratio of coronal integrated total wave energy 
to chromospheric integrated total wave energy (solid lines) and the ratio of coronal integrated 
Poynting flux to chromospheric integrated Poynting flux (dashed lines). The two lines for each 
quantity correspond to the maximum and minimum ratios possible, reflecting the uncertainty in known 
values of plasma parameters.  }\label{fig:damp}
\end{figure}
  
\subsection{Wave damping}
Let us now compare the chromospheric power spectrum to those derived for the corona 
(\citealp{TOMMCI2009}). The underlying assumption has to be made that the coronal observations taken 
by the Coronal Multi-channel Polarimeter (CoMP) and the ROSA observations both give results typical of 
quiescent Sun phenomena for the corona and chromosphere, respectively. We note the CoMP data is also 
of a quiescent Sun region. Large, faint loop structures are barely visible in CoMP and Solar and 
Heliospheric Observatory 195~{\AA} images. Examining the region as it rotates onto the disk reveals no 
visible signs of large magnetic flux concentrations in Solar and Heliospheric Observatory magnetograms. 
Hence, we assume the region is also typical of the quiescent Sun. However, we note that the general 
magnetic topology throughout the solar corona could well be different between the two datasets. This is 
because the ROSA data is taken further into the solar cycle than the CoMP data.

Figure~\ref{fig:damp}a shows the velocity power measured with CoMP for a coronal loop arcade 
structure divided by the fitted power law for the 
chromospheric velocity power from region D2, for the period range 100-500~s. The ratio reveals that the 
velocity power appears to decrease significantly from the chromosphere to the corona, with the power of 
the high-frequency waves decreasing to a much greater degree. The velocity power of the CoMP data is, 
however, averaged over a distance of 250~Mm along a coronal loop system, where frequency-dependent 
wave damping has occurred (\citealp{TOMMCI2009}; \citealp{VERTHetal2010}).  However, for our purpose 
it is necessary to calculate the input velocity power at the base of the coronal loop system. The minimum 
height in the solar atmosphere that CoMP can measure Doppler velocities is $\sim20$~Mm above the 
solar surface. The input power spectra at the base of the CoMP loop system (at a height of 20 Mm) can be 
determined by exploiting the measured damping rates (\citealp{VERTHetal2010}). 

First, the spatially averaged total power as a function of frequency, $f$, is denoted by $\langle P(f)\rangle_{Total}$. 
This averaged velocity power detected by CoMP is composed of waves 
propagating both upwards and downwards along the loop path. The particular averaged velocity power of 
the waves that propagate from the loop base, $s=0$ to the loop apex, $s=L$, is denoted $\langle P(f)\rangle_{up}$. 
The averaged velocity power of waves propagating downwards along the loop path is denoted
$\langle P(f)\rangle_{down}$. These latter waves will have been generated at the other loop foot-point so once the 
reach the apex they will have already travelled a distance $L$. The spatially averaged total power is then 
given by
\begin{equation}
\langle P(f)\rangle_{Total}=\langle P(f)\rangle_{up}+\langle P(f)\rangle_{down}.
\end{equation}
Now, modelling the observed damping (\citealp{TOMMCI2009}) of the waves as they propagate between 
the loop base,
$s=0$, and the loop apex, $s=L$, the averaged total velocity power is given by,
\begin{eqnarray}
\langle P(f)\rangle_{Total}&=&\frac{1}{L}\int_0^{L}P(f)_{in}\exp\left(-\frac{2s}{L_D}\right)\mathrm{d} s\nonumber\\
&&+\frac{1}
{L}\int_L^{2L}P(f)_{in}\exp\left(-\frac{2s}{L_D}\right)\mathrm{d} s,
\end{eqnarray}
where $P_{in}(f)$ is the power input at the CoMP base height (i.e., 20~Mm), $L_D=(\tau/P)(c_k/f)$ is the
damping length, $c_k$ is the kink wave phase speed and $\tau/P$ is the quality factor 
(\citealp{VERTHetal2010}). Integrating and re-arranging gives
\begin{equation}\label{eq:input}
P(f)_{in}=\frac{2L}{L_D}\langle P(f)\rangle_{Total}\left[1-\exp\left(-\frac{4L}{L_D}\right)\right]^{-1}.
\end{equation}
Supplementing Equation~\ref{eq:input} with $c_k=0.6$~Mm s$^{-1}$ (\citealp{TOMMCI2009}), 
$\tau/P=2.69$ (\citealp{VERTHetal2010}) and $L=250$~Mm, we determine the dashed lines in 
Figure~\ref{fig:damp}a.

The gradient for the frequency dependent trend is now shallower but still gives about factor of five 
decrease in velocity power for the higher frequency waves ($f\approx10$~mHz) relative to the 
lower frequency waves ($f\approx2$~mHz). This behaviour is also shown for the transmission profile of 
the velocity amplitude in Figure~\ref{fig:damp}b, where the decrease in velocity amplitude is about a 
factor of three over this frequency range.

 We now determine the energy loss between the chromosphere and corona. The following relations show 
the time-averaged, spatially integrated total energy ($E$) and Poynting flux ($S_z$) for MHD kink waves 
(\citealp{GOOetal2013}),
\begin{equation}
E=(\rho_i+\rho_e)v_r^2\pi R^2,\qquad S_z=(\rho_i+\rho_e)v_r^2c_k\pi R^2
\end{equation}
where
\begin{equation}
c_{k}^2=\frac{B_i^2+B_e^2}{\mu_0(\rho_i+\rho_e)}
\end{equation}
is the phase (propagation) speed of the wave, $\rho$ is the density, $v_{r}$ is the velocity perturbation, $\mu_0$ is the magnetic permeability of free space and $R$ is the radius of the flux tube. The
subscripts refer to the internal, $i$, and ambient, $e$, plasma quantities.

Due to significant uncertainties in the values of the equilibrium plasma 
parameters, we calculate the minimum and maximum ratios possible for both quantities.
In general, estimates of density are $10^{-12}-10^{-13}$~kg\,m$^{-3}$ in the corona
(\citealp{WARBRO2009}) and $10^{-9}-10^{-11}$~kg\,m$^{-3}$ for the chromosphere 
(\citealp{BEC1968}). The measured radii of flux tubes in the chromosphere are 100-400~km 
(\citealp{MORetal2012c}), while measurements from TRACE (\citealp{WATKLI2000}) and Hi-C 
(\citealp{MORMCL2013}) suggest  100-1000 km for coronal loops. Finally, measured phase speeds of 
kink waves in the chromosphere suggest the waves propagate at 100-250 km\,s$^{-1}$ 
(\citealp{MORetal2012c}, \citealp{OKADEP2011}] and $600-1000$~km\,s$^{-1}$ in the corona 
(\citealp{TOMMCI2009}). The parameters of density, magnetic 
field, phase speed and tube radius are height dependent. Physically this means the values have to satisfy
the following relations,
$$
\frac{c_{k1}}{c_{k2}}=\sqrt{\frac{\rho_2(1+\chi_2)}{\rho_1(1+\chi_1)}}\frac{B_1}{B_2}, \qquad
\frac{R_1}{R_2}=\sqrt{\frac{B_2}{B_1}},$$
$$
 \qquad B_1\leq B_2, \qquad \rho_1\leq\rho_2.
$$
Here the subscripts $1, 2$ refer to the coronal and chromospheric values respectively,
$\chi=\rho_e/\rho_i=0.1-0.5$ and we assume $B_i\approx B_e$. The second of these equations is derived from
the conservation of magnetic flux, the third assumes that the magnetic field strength does not increase
with height; the fourth assumes density does not increase with height.

In Figure~\ref{fig:damp}c, the ratio of the integrated total energy and integrated Poynting flux between 
the corona and the chromosphere are shown.  

Figure~\ref{fig:damp}c 
demonstrates that even the upper bound of the ratio suggests only a transmission of 0.01\% of the total 
chromospheric kink wave energy and Poynting flux. These findings come with the following caveat: the 
spatial sampling ($\approx4.5$~Mm) of CoMP means it may not resolve the coronal fine structure adequately. It 
has been demonstrated the effect of LOS integration on multiple unresolved loops leads to an 
underestimate of kink wave velocity amplitude (\citealp{DEMPAS2012}; \citealp{DEPMCI2012}). At 
present, we cannot give an explicit value of CoMP’s under-resolution; therefore, Figure~\ref{fig:damp} 
represents the transmission profiles of the velocity power, amplitude, etc., rather than the absolute 
values of the transmission coefficients. {We point out that non-thermal widths from the CoMP observations
are much larger ($30-40$~km/s) the the Doppler shifts and may  provide a better indicator of the amplitudes of the
unresolved kink motions. However, the non-thermal widths are likely to also include contributions from the line-of-sight
components of torsional motions, flows, slow waves, magneto-acoustic waves, etc., which complicates their interpretation. }


\begin{figure*}[!htp]
\centering
\includegraphics[scale=0.8, clip=true, viewport=0.0cm 0.0cm 14.5cm 12.0cm]{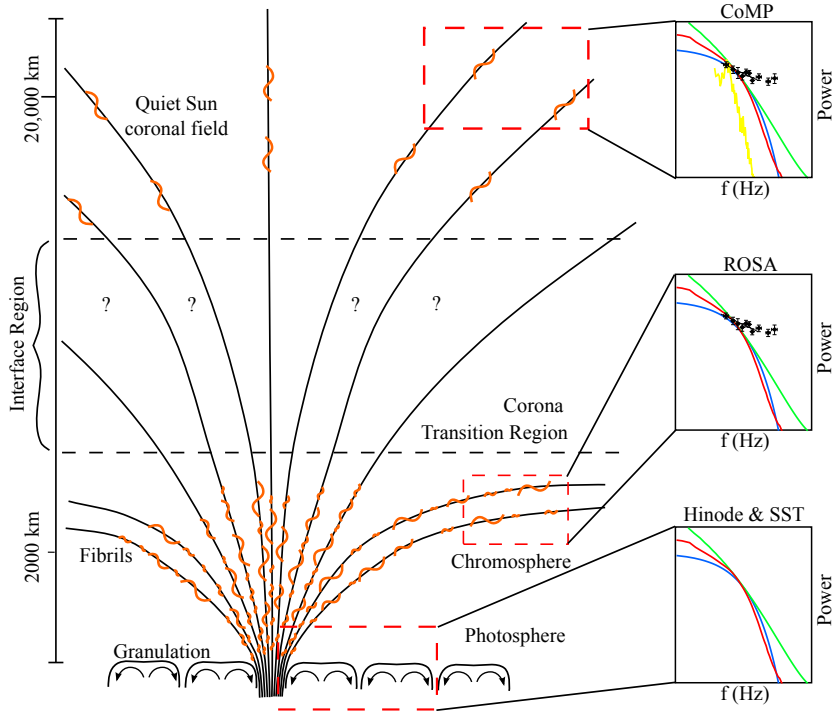} 
\caption{The transport of energy through the quiescent solar atmosphere. A cartoon depicting the 
transport of kink MHD wave energy through the solar atmosphere as implied by observational results. 
The kink MHD waves are assumed generated by the horizontal granular motions, which imparts a 
particular power spectra on the waves. Kink MHD waves observed in the quiescent chromosphere 
demonstrate a similar power spectrum suggesting the granular motions have indeed excited them. The 
waves then enter a region that is difficult to observe with to current solar instrumentation. We refer to 
this as the interface region and it consists of the Transition Region and low Corona. On reaching the 
upper Corona, CoMP measurements reveal that there has been a significant loss of higher frequency wave 
energy between the chromosphere and Corona. It could well be that the energy has been dissipated and 
is heating the solar atmosphere.
\vspace{0.2cm}
}\label{fig:cart}
\end{figure*}

The observed frequency-dependent trends of the quantities in Figure~\ref{fig:damp} can be explained by 
frequency-dependent damping, which has been well studied in the case 
of kink wave damping in coronal loops (\citealp{VERTHetal2010}; \citealp{VERetal2013}). From 
observations of such damped standing kink waves the ratio of damping time ($\tau$) over period 
provides the relative strength of the damping mechanism, i.e., the quality factor. Statistical studies show 
coronal values of $\tau/P\approx 1-5$, meaning the observed kink waves lie in the under-damped 
regime (\citealp{VERetal2013}). In the case of propagating waves the equivalent measure is 
$L_D/\lambda$, where $L_D$ is the damping length and $\lambda$ is the wavelength 
(\citealp{TERetal2010c}). Analysis of damped propagating kink waves detected in the CoMP data reveal 
that the two measures are in agreement (\citealp{VERTHetal2010}), i.e., $\tau/P\sim L_D/\lambda$. 
Assuming that frequency-dependent damping also occurs in the interface region between ROSA and 
CoMP observations, with $L_D =(\tau/P) c_k/f$, where $c_k$ is the kink wave phase speed, we can 
estimate the interface region damping lengths using the following relation
\begin{equation}
P_{out, \,IR}(f)\propto P_{in, \,IR}(f)\exp\left(-\frac{2L}{L_D}\right),
\end{equation}
where $P_{out, \,IR}(f)$ is the output power from the interface region to the corona (i.e., the CoMP 
footpoint power spectrum) and $P_{in, \,IR}(f)$ is the input power from the chromosphere to the interface 
region (i.e., the ROSA power spectrum). Writing $L_D=\alpha/f$, where $\alpha=(\tau/P)c_k$, and using 
estimates of the height of the interface region, i.e., L=15-20~Mm, we find $\alpha_{IR}\approx0.2$~Mm/s 
provides a reasonable approximation of the  change in gradient between ROSA and CoMP velocity power 
spectra.

From the analysis of the damping of propagating coronal kink waves $\alpha_{corona}\approx1.6$~Mm/s 
(\citealp{VERTHetal2010}), it follows that the damping length in the interface region is about $13\%$ of 
the estimated coronal damping length. The quality factors in the interface region and corona are related 
by the following equation,
\begin{equation}
\left(\frac{\tau}{P}\right)_{IR}=\left(\frac{\alpha}{c_k}\right)_{IR}\left(\frac{c_k}
{\alpha}\right)_{corona}\left(\frac{\tau}{P}\right)_{corona}.
\label{QF}
\end{equation}
To estimate $(\tau/P)_{IR}$ from Equation~(\ref{QF}) the previous values are used for $\alpha_{IR}$, 
$\alpha_{corona}$, $c_{k, corona}$, $c_{k, \, IR }\approx150$ km s$^{-1}$ is taken for the average 
phase speed in the interface region (between heights of 5 and 20 Mm - \citealp{MCIetal2011}), and $({\tau}/{P})_{corona}=2.69$. This gives 
$(\tau/P)_{IR} \approx 1.35$, about 50\% of the coronal value. Hence, the observed trend implies that 
there is much stronger frequency-dependent kink wave damping in the lower solar corona. Critical 
damping occurs when 
$(\tau/P)=1/(2 \pi)\approx 0.16$, so although the estimated quality factor in the interface region is 
about a half of that in the corona, it is still above the critical damping regime. This is consistent with
the fact that these propagating waves are actually observed at higher altitudes with CoMP, i.e., they are 
not completely killed off within the interface region. The intimation of enhanced and 
frequency-dependent kink wave damping between the chromosphere and corona has potentially important 
implications for numerous coronal-heating models, which demonstrate that incompressible wave energy 
is more efficiently converted to heat at higher frequencies. Furthermore, the particular location of 
enhanced kink wave damping is significant because there is mounting observational evidence for the 
chromosphere and interface region to be the predominant locations for plasma heating processes in the 
solar atmosphere (\citealp{ASCetal2007}; \citealp{TRIetal2012}).

\bigskip

The results presented here show that the measurement of velocity power spectra provides a very 
powerful and practical mechanism for analysing MHD kink wave propagation through the magnetised 
solar atmosphere. Comparing the velocity power spectra obtained at different altitudes of the atmosphere 
allows for the possible signatures of kink wave driving and damping to be observed 
(Figure~\ref{fig:cart}). The picture implied by the observations presented here suggests a qualitative agreement with theoretical 
expectations for wave propagation through the quiescent solar atmosphere (e.g., \citealp{CHOetal1993}, \citealp{CRAVAN2005}), i.e., 
magnetic waves are driven by the horizontal motions that propagate into the upper solar atmosphere, with the flow of wave energy 
hindered by the strong gradients present in the transition region. 

These observational results do not tell the complete picture though and they raise a number of key questions 
that need to be answered, e.g., what is the fate of the high-frequency wave energy observed in the 
chromosphere?; What mechanism(s) has led to their decrease in power before they reach the 
corona? One possible explanation of coronal kink wave damping is through mode conversion to m=1 
torsional Alfv\'en waves at resonant magnetic surfaces naturally present across inhomogeneous solar 
atmospheric waveguides. Such a process could also explain the stronger kink wave damping in the 
interface layer (the 15-20 Mm region linking between ROSA and CoMP observations). In the lower 
atmosphere (at heights of less than 10 Mm) it has recently been shown that torsional Alfv\'en and kink 
waves are concurrent in spicules (\citealp{DEPetal2012}), providing evidence that mode coupling is 
already happening at sub-interface region heights. To fully understand the interaction and evolution of 
these coupled incompressible MHD wave modes in the interface layer between the chromosphere and 
corona, missions such as the Interface Region Imaging Spectrometer will be crucial.

\acknowledgements{RE acknowledges M. Kéray for patient encouragement. AH is supported by the 
Grant-in-Aid for Young Scientists (B, 25800108). The authors are also grateful to NSF, Hungary (OTKA, 
Ref. No. K83133), the Science and Technology Facilities Council (STFC), UK. GV acknowledges the Leverhulme Trust. RM 
acknowledges Northumbria University for the award of the Anniversary Research Fellowship and the Royal Astronomical Society for the 
award of travel grants. Observations were obtained at the National Solar Observatory, operated by the Association of Universities for 
Research in Astronomy, Inc. (AURA), under agreement with the National Science Foundation. We would like to thank the technical staff at 
DST for their help and support during the observations. Further thanks are required for  M. Mathioudakis, D. B. Jess and Queen's 
University Belfast, UK for ROSA support and S. Tomczyk, T. Matsumoto and L. P. Chitta who provided us with the results from other 
instruments. }

\bibliographystyle{aa}

\end{document}